\documentclass[slac_one]{revtex4}

\usepackage{graphicx}
\usepackage{fancyhdr}
\usepackage{subfigure}

\begin{document}

\title{Generalization of exotic quark searches}

\author{F. Garberson, T. Golling}
\affiliation{Yale University, New Haven CT, 06520, USA}

\begin{abstract}

  General limits on exotic heavy quarks $T$, $B$ and $X$ with masses
  above 300~GeV are presented for arbitrary branching fractions of
  $T\to W^+b$, $T\to Zt$, $T\to Ht$, $ B\to W^-t$, $B\to Zb$, $B\to
  Hb$ and $X\to W^+t$.  The results are based on a CMS
  search in final states with three isolated leptons ($e$ or $\mu$)
  or two isolated leptons with the same electric charge. Exotic heavy
  quark pair production through the strong interaction is considered.
  In the context of vector-like quark models, $T$ quarks with a mass
  $m_T < 480$~GeV and $m_T < 550$~GeV are excluded for weak isospin
  singlets and doublets, respectively, and $B$ quarks with a mass $m_B
  < 480$~GeV are excluded for singlets, all at 95\% confidence level.
  Mass limits at 95\% confidence level for $T$ and $B$ singlets,
  ($T$,$B$) doublets and ($X$,$T$) doublets are presented as a
  function of the corresponding heavy quark masses.  For equal mass
  $m_T = m_B$ and $m_X = m_T$ vector-like quarks are excluded at 95\%
  confidence level with masses below 550~GeV for $T$ and $B$ singlets,
  640~GeV for a ($T$,$B$) doublet and 640~GeV for a ($X$,$T$) doublet.

\end{abstract}

\maketitle

\pagestyle{plain}



\section{\label{sec:introduction}Introduction}

Vector-like quarks~\cite{AguilarSaavedra:2009es, bib:SingletSaavedra, bib:AguilaExoticQuark} are new heavy quarks,
in particular heavier than the top quark, which appear in many new
physics models, such as extra dimensions, little Higgs, or composite Higgs
models.  Similar to a supersymmetric partner of the top quark, a
vector-like top partner serves to stabilize the Higgs mass by
cancelling the divergence of radiative corrections in the Higgs boson
mass.  Quarks are referred to as vector-like if their left- and
right-handed chiralities transform in the same way under the
electroweak group $SU(2) \times U(1)$.  Vector-like quarks can be
classified as weak isospin singlets, doublets or triplets.  The mass
eigenstates of these vector-like quarks are referred to as $T$ and $B$,
with charges $2/3$ and $-1/3$, respectively, and $X$ and $Y$, with
charges $5/3$ and $-4/3$, respectively.  It is assumed that the new
quarks mainly couple to the third generation~\cite{delAguila:1982fs}
which leads to the following possible decay modes:

$$ T\to W^+b, \;\;\; T\to Zt, \;\;\; T\to Ht,$$
$$ B\to W^-t, \;\;\; B\to Zb, \;\;\; B\to Hb,$$
$$ X\to W^+t, $$
$$ Y\to W^-b. $$

For $T$ and $B$ singlets all decay modes are sizable. For
doublets a reasonable assumption is that $V_{Tb}\ll V_{tB}$ so that 
only $T\to Zt$, $T\to Ht$ and $B\to W^-t$ contribute. In this paper we 
give special attention to this scenario but also present results that 
can be interpreted under more general CKM hypotheses.
 In addition, the branching
fractions for the $T$ and $B$ decay modes vary with the heavy quark
masses $m_T$ and $m_B$, respectively.  The total width of the new
quarks is typically negligible as compared to the detector mass
resolution in the probed mass range.

Exotic heavy quarks such as vector-like quarks are mainly produced in
pairs through the strong interaction or singly via the electroweak
interaction.  We will only focus on the pair production.  The cross
section for this process is the same for each of these types of quarks
and depends on the quark mass.

Using 4.9~fb$^{-1}$ of $pp$ collision data at $ \sqrt{s} = 7$~TeV the
CMS Collaboration excluded the existence of a fourth-generation $b'$
quark with a mass below 611~GeV at 95\% confidence level
(CL)~\cite{Chatrchyan:2012yea} by examining events with three isolated
leptons ($e$ or $\mu$) or two isolated leptons with same electric
charge.  The ATLAS collaboration used a same-sign dilepton data
sample equivalent to 4.7~fb$^{-1}$ of $pp$ collisions at $\sqrt{s} =
7$~TeV to exclude both $b'$ and vector-like quarks $X$ with charge
$5/3$ (which they referred to as $T_{5/3}$) with masses below 670~GeV at 95\%
CL~\cite{bib:ATLASSSres}.  A similar analysis was performed by the CMS
Collaboration using 5~fb$^{-1}$ of $pp$ collisions at $\sqrt{s} =
7$~TeV to exclude vector-like quarks $X$ with charge $5/3$ with masses
below 645~GeV at 95\% CL~\cite{CMS-PAS-B2G-12-003}.  All three
searches assume pair production through the strong interaction and
branching ratios of unity $BR(b' \to W^-t) = 1$ and $BR(X \to W^+t) =
1$.

However, same-sign-lepton or three-lepton signatures are expected from
many of the vector-like-quark final states discussed above.
Accounting for all possible decay processes, for $T\bar{T}$ production
the possible final states are $W^+bW^-b$, $WbZt$, $WbHt$, $ZtZt$,
$ZtHt$ and $HtHt$. For $B\bar{B}$ production the possible final states
are $W^-tW^+t$, $WtZb$, $WtHb$, $ZbZb$, $ZbHb$ and $HbHb$. For
$X\bar{X}$ production the only possible final state is $W^+tW^-t$. For
simplicity, here $b$ represents both $b$ and $\bar{b}$, and
analogously for top quarks.  Among these, all final states except
$W^+bW^-b$ feature same-sign-lepton or three-lepton signatures.  In
this paper we exploit this feature and reinterpret the published CMS
result~\cite{Chatrchyan:2012yea} by relaxing the assumptions which
determine the branching ratios as a function of mass to explore the
entire space of possible branching ratios.  A similar reinterpretation
for arbitrary branching fractions was performed by
ATLAS~\cite{ATLAS:2012qe} for an analysis targeting the $T\bar{T} \to
W^+bW^-b$ hypothesis in a single-lepton final state and excluding at
95\% CL $T$ quarks with a mass 400~GeV~$< m_T < 500$~GeV for weak
isospin singlets.  We show that these limits can be extended with the
same-sign-lepton and three-lepton signatures, and we present limits
for the $T$ and $B$ singlet and doublet, as well as for the ($X$,$T$)
doublet hypotheses as a function of the corresponding heavy quark
masses.  These represent important generalizations of the existing
limits.

\section{\label{sec:samples_and_selection}Samples and event selection}

In this analysis the event selection of the CMS search is replicated
as closely as possible. The selection is briefly described in
Section~\ref{sec:ExperimentSel}. The details of how the selection is
reproduced for this paper are discussed in Section~\ref{sec:OurSel}.

\subsection{\label{sec:ExperimentSel} The CMS event selection}

As discussed above, CMS makes use of events that are selected under
both same-sign and trilepton requirements (electron or muon). Events
are selected if they pass a trigger that requires two
leptons. Electrons are required to have $p_{T} > 20$ GeV and $|\eta| <
2.4$, excluding the region between the end-cap and barrel ($1.44 <
|\eta| < 1.57$). Muons are required to have $p_{T} > 20$ GeV and
$|\eta| < 2.4$. Jets are required to have $p_{T} > 25$ GeV and $|\eta|
< 2.4$. For the same-sign analysis, CMS requires the presence of two
isolated leptons with the same electric charge and at least four
jets. For the trilepton analysis at least three isolated leptons must
be identified, at least two of which must have an opposite charge, and
at least two jets must be found. In all cases, at least one jet must
be tagged as a $b$-jet using a tagger with roughly a 50\% efficiency
for identifying true $b$-jets, and events with two electrons or muons
that are consistent with originating from a $Z$ boson decay are
rejected ($|m_{ll} - m_Z| > 10$ GeV). Finally, the scalar sum of the
transverse momenta of the jets, leptons, and the missing transverse
momentum is required to be at least 500~GeV.


\subsection{\label{sec:OurSel} Samples and event selection for the reinterpretation}
For this analysis samples of singlet $T\bar{T}$ and $B\bar{B}$
production, and doublet $X\bar{X}$ production were generated. All
samples were generated using Protos
2.0~\cite{AguilarSaavedra:2009es,bib:Protos} and showered using Pythia
6.4.25~\cite{bib:Pythia6}. 500,000 events were generated for each
process and each mass hypothesis in 50 GeV intervals ranging from a
lowest mass of 300~GeV to a highest mass of 900~GeV. In each case the
Higgs mass was set to a value of 125 GeV.

The modeling of the CMS detector was performed using the Pretty Good
Simulation (PGS) package~\cite{bib:PGS}, with the detailed detector
descriptions taken from the default CMS detector card from the
MadGraph~\cite{bib:MadGraph} package. A few minor changes were then
made to these defaults in order to improve the accuracy as discussed
below.  Some information, such as the CMS efficiencies for $b$-tagging
and electron identification, is not provided in precise detail by the
experiment. We document our assumptions below, and will show that we
achieve good agreement with the published results.

All kinematic cuts on $p_{T}$\ and $\eta$\ of the leptons and jets are
chosen to be identical to those of the experiment. PGS applies its own
model for electron identification but it does not attempt to determine
efficiency loss due to isolation cuts. In order to make the model more
realistic we remove electrons if they are found to be within $\Delta R
< 0.4$ of a jet with $p_{T} > 15$ GeV unless $\Delta R < 0.2$, in
which case the jet is removed instead under the assumption that it is
a misidentified electron.
As the CMS documentation~\cite{bib:CMSElectron} does not give 
precise numbers for the selection efficiencies, we assume that their
electrons have the same isolation efficiencies as for
ATLAS~\cite{bib:ATLASSSres}.  The calorimeter and tracker cuts are
each 90\% efficient for the electron isolation at ATLAS.
  
Unlike for electrons, for muons PGS does not have a built-in muon
identification model. Instead it only assumes a 2\% inefficiency on
all tracks in the analysis. This efficiency appears to be roughly
correct for CMS, which has highly efficient muon selection and
isolation requirements~\cite{bib:CMSMuon}. 

CMS has triggers that are highly efficient for
electron identification, but less efficient for muon
identification. CMS explicitly quotes their trigger efficiencies for
their selected events as being 91\% in the $\mu\mu$ channel, 96\% in
the $e\mu$\ channel, and 99\% in the $ee$\
channel~\cite{Chatrchyan:2012yea}. When simulating the same-sign
dilepton analysis for CMS, events are randomly thrown out according to
these probabilities. For the CMS trilepton analysis the triggers are
assumed to be 100\% efficient.  

The $b$-tagging efficiency for state-of-the-art CMS 
$b$-tagging algorithms is quite different from the efficiency that is
assumed by PGS. We therefore set the $b$-tagging efficiency to more
appropriate values. 
For CMS a tagger
was chosen that was tuned to be 50\% efficient for real $b$-jets with
a 1\% mistagging efficiency for non-$b$
jets~\cite{Chatrchyan:2012yea}. In principle it would be best to 
account for the $p_{T}$ and $\eta$-dependence of the tagging
efficiencies. Unfortunately, none of the CMS public $b$-tagging
documents~\cite{bib:CMSbTagEff,bib:CMSbTagEff2} provide the efficiencies of this
particular operating point as a function of jet kinematics. This
documentation does, however, indicate that the $b$-taggers in CMS tend
to have less kinematic dependence than at many other experiments. We therefore
instead use the average efficiency of 50\% for real $b$-jets and 1\%
for non-$b$ jets as quoted in the paper~\cite{Chatrchyan:2012yea}. 
%

After applying all event selection, the PGS event yields are validated
against the quoted CMS yields for the pair production of $b'$ quarks.
As shown in the ATLAS analysis~\cite{bib:ATLASSSres}, these yields are
approximately identical to the $X$ pair production yields.  For purposes of this
validation, since CMS
assumes $BR(b' \to W^-t) = 1$~\cite{Chatrchyan:2012yea,bib:ATLASSSres,CMS-PAS-B2G-12-003},
we apply a filter to force 
the vector-like-quarks to decay in the same manner ($BR(B \to W^-t) = 1$).
Very good agreement is found
as shown in Table~\ref{tab:CMSCombSelComp}.


\begin{table}[th]
  \begin{center}
  \begin{tabular}{|l|c|c|c|c|c|}
\hline 
 	&	  $m_{b'/B}$ = 450 GeV & $m_{b'/B}$ = 500 GeV & $m_{b'/B}$ = 550 GeV & $m_{b'/B}$ = 600 GeV & $m_{b'/B}$ = 650 GeV \\
\hline 
PGS same-sign  		&	 49.9 $\pm$ 1.4 & 25.1 $\pm$ 0.7 & 14 $\pm$ 0.4 & 7.9 $\pm$ 0.2 & 4.1 $\pm$ 0.1 \\
CMS same-sign  	&	 49 $\pm$ 4.2 & 26 $\pm$ 2.2 & 14 $\pm$ 1.1 & 7.6 $\pm$ 0.6 &  4.3 $\pm$ 0.4 \\
CMS SS Difference (\%) 	&	 1.8 $\pm$ 9 & -3.4 $\pm$ 8.9 & 0 $\pm$ 8.6 & 3.7 $\pm$ 8.7 & -4.1 $\pm$ 9.1 \\
\hline 
PGS trilepton  	&	 16 $\pm$ 0.8 & 8.1 $\pm$ 0.4 & 4.5 $\pm$ 0.2 &  2.6 $\pm$ 0.1 & 1.5 $\pm$ 0.1 \\
CMS trilepton  	&	 15 $\pm$ 1.6 & 8.2 $\pm$ 0.8 & 4.7 $\pm$ 0.4 & 2.7 $\pm$ 0.3 & 1.6 $\pm$ 0.2 \\
CMS trilepton Difference (\%) 	&	 6.6 $\pm$ 11.9 & -0.6 $\pm$ 10.9 & -3.5 $\pm$ 9.9 & -5.3 $\pm$ 10.8 & -8.3 $\pm$ 10.2 \\
\hline 
  \end{tabular}
  \caption{Comparison of the number of expected signal events passing
    all selection requirements between PGS and the quoted results from the CMS paper 
    for both the same-sign and the trilepton selections, normalized
    to the CMS integrated luminosity of 4.9~fb$^{-1}$.    Uncertainties
    on the CMS paper results include both statistical and systematic
    uncertainties, while uncertainties on the PGS result are statistical only. } \label{tab:CMSCombSelComp}
  \end{center}
\end{table}





\section{\label{sec:method}Method}

This analysis proceeds in three steps. First, events are selected
according to the prescriptions documented in
Section~\ref{sec:samples_and_selection}. It must be remembered that in
the case of $B\bar{B}$ and $T\bar{T}$ events, the simulation assumes
that the heavy quarks are singlets. When predictions for a model with
alternate decays such as a doublet model are desired instead, a
correction is needed to the appropriate decay branching
fractions. This procedure is described in
Section~\ref{sec:BRConversion}. Finally, the number of observed events
is converted into exclusion limits. This procedure is discussed in
Section~\ref{sec:LimitSetting}.

%
%

\subsection{\label{sec:BRConversion} Alternate decay-mode hypotheses}

Depending on how each of the new heavy quarks decays, there are six
possible combinations for each quark type as explained in
Section~\ref{sec:introduction}. The nominal $T\bar{T}$ and $B\bar{B}$
samples in this analysis are generated with the hypothesized branching
fractions for the new heavy quark decays that is appropriate for
singlets. In this section we explain how the expected number of signal
events is converted to a value that is appropriate to more general
branching fractions such as under a doublet model.

To achieve full generality, for each mass point a two dimensional grid
of all possible branching fractions is scanned in 10\% steps. The
dimensions are chosen to be the branching ratio of the $W$-type decay
modes, which we denote $B_{W}$, and the branching ratio of the
$Z$-type decay modes, which we denote $B_{Z}$. The branching ratio of
the Higgs-type decays follows from $B_{H} = 1 - B_{W} - B_{Z}$. The
probability for the production of each of the six possible
combinations of decays of the two new heavy quarks then depends upon
these branching fractions that we are assuming. For a particular decay
mode $i$, the probability is denoted $P_{i}(B_{W}, B_{Z})$. After
determining the acceptance times efficiency for our event selection
for each decay, $A_i$, the number of signal events $N$ that is
expected for each hypothesis branching fraction is then determined
according to Equation~\ref{Eq:BR}:

\begin{equation}  \label{Eq:BR}
N = \sum^{6}_{i=1} P_{i}(B_{W}, B_{Z}) A_{i} \int \mathcal{L} \sigma,
\end{equation}

where $\int \mathcal{L}$ is the integrated luminosity and $\sigma$ is
the corresponding heavy quark pair production cross section.  We
assume that the kinematic differences and consequently the differences
in selection efficiency for singlets and doublets are negligible.

\subsection{\label{sec:LimitSetting} Limit setting}

In order to set limits on a given model it is necessary to know the
following information for each measurement: the number of expected
background events from each source with their associated
uncertainties, the number of expected signal events with their
associated uncertainty, and the number of observed data events.

The number of data events, the number of
background events from each source, and their respective uncertainties
are taken directly from the CMS paper. The predictions for
the signal model are taken from the outputs of the PGS simulation,
with any necessary corrections applied as discussed in
Sections~\ref{sec:OurSel} and \ref{sec:BRConversion} to arrive at a
given decay hypothesis. In all cases, systematic uncertainties on the
signal are assumed to have the same relative size as are quoted in the paper.

Generally it is necessary to split the uncertainties for each sample
into their components in order to correctly handle correlations
between the signal and background samples.
However, in all cases considered here, the systematic uncertainties
that are correlated between samples are negligibly small and can be
safely neglected. In particular, for the same-sign analysis the
dominant background systematics are due to control-region estimations,
which are not correlated to the signal uncertainties. Similarly, in the trilepton
analysis the dominant background uncertainties are from data statistics, 
normalization of the theoretical backgrounds, and Monte Carlo sample statistics, 
which again are not correlated to the signal uncertainties. We therefore
neglect correlations when running the limit-setting. In tests we were able to
reproduce the limit results for the CMS results (and for the similar ATLAS 
results~\cite{bib:ATLASSSres}) to within 10 GeV in mass under this assumption.

\begin{figure*}
\centering
    \includegraphics[height=3.5in]{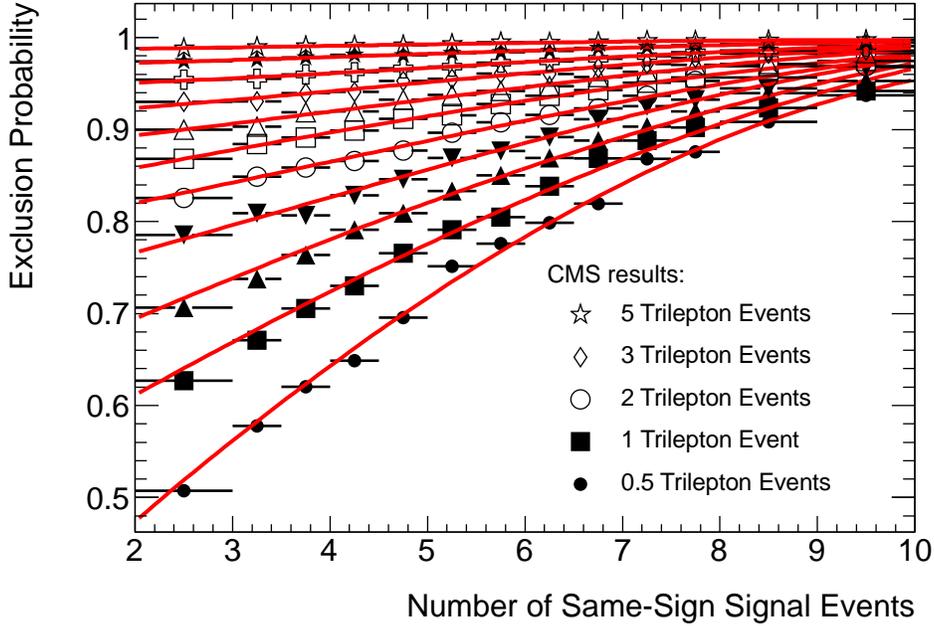}
\caption{The expected exclusion probabilities as a function of the
  number of expected signal events are plotted. 
A two-dimensional parametrization is needed in the predicted
  number of same-sign and trilepton events. The $x$-axis 
  shows the number of same-sign signal events, while fit functions are
  overlaid depending on the number of trilepton signal events. Eleven
  functions are shown for bins of between 0.5 and 6.0 trilepton
  events, representing bins with the following bounds: $\{0.5, 1.0,
  1.33, 1.67, 2.0, 2.33, 2.67, 3.0, 3.5, 4.0, 5.0, 6.0\}$. The goal is
  to determine where each parametrization crosses the 95\% threshold
  exclusion probability. A linear interpolation is applied to determine 
  the proper functional form for
  the number of trilepton events in between bin centers.}
\label{fig:LimitParam}
\end{figure*}

Limit setting is performed by running the
MCLimit~\cite{bib:LimitTJunkArXiv, bib:LimitAReadJPhysG} program
simultaneously on the same-sign and the trilepton results. In order
for the limit setting to converge with high accuracy, tens (or
hundreds) of thousands of pseudo-experiments must be performed for
each signal hypothesis. When scanning over all of the mass and
branching-fraction parameters that are considered for the signal
models in this paper this procedure becomes very computationally
intensive. A simplification is therefore made in order to streamline
the process. Since results of the limit setting will depend only on
the predicted signal yields for each model, the exclusion probability
can therefore be parametrized in two variables: the number of
predicted same-sign and trilepton signal events. These variables are 
scanned and MCLimit predictions are made at periodic steps. Functions
are fit to interpolate and smooth the results between these
points. Results are shown in Figure~\ref{fig:LimitParam}.


\section{\label{sec:results}Results}
In Section~\ref{sec:AlternateBR} the results of the analysis assuming
arbitrary branching fractions for heavy quark $T\bar{T}$ and
$B\bar{B}$ decays are presented. In Section~\ref{sec:NominalBR} these
results are interpreted in the context of certain theoretically
motivated values of heavy quark branching fractions, including the
$(X,T)$ doublet.

\subsection{\label{sec:AlternateBR} Results for arbitrary branching
  fractions for the new heavy quark decays}

In this section results are presented for all possible branching
fractions of heavy $T$ and $B$ quark decays. In each case we assume
only the presence of a single $T\bar{T}$ or $B\bar{B}$ production
process. Having both $T\bar{T}$ or $B\bar{B}$ present would of course
lead to improved sensitivity. We discuss some such models with two new 
heavy quarks later in Section~\ref{sec:NominalBR}.

\begin{figure*}
\centering
\subfigure[] {
  \includegraphics[height=2.3in]{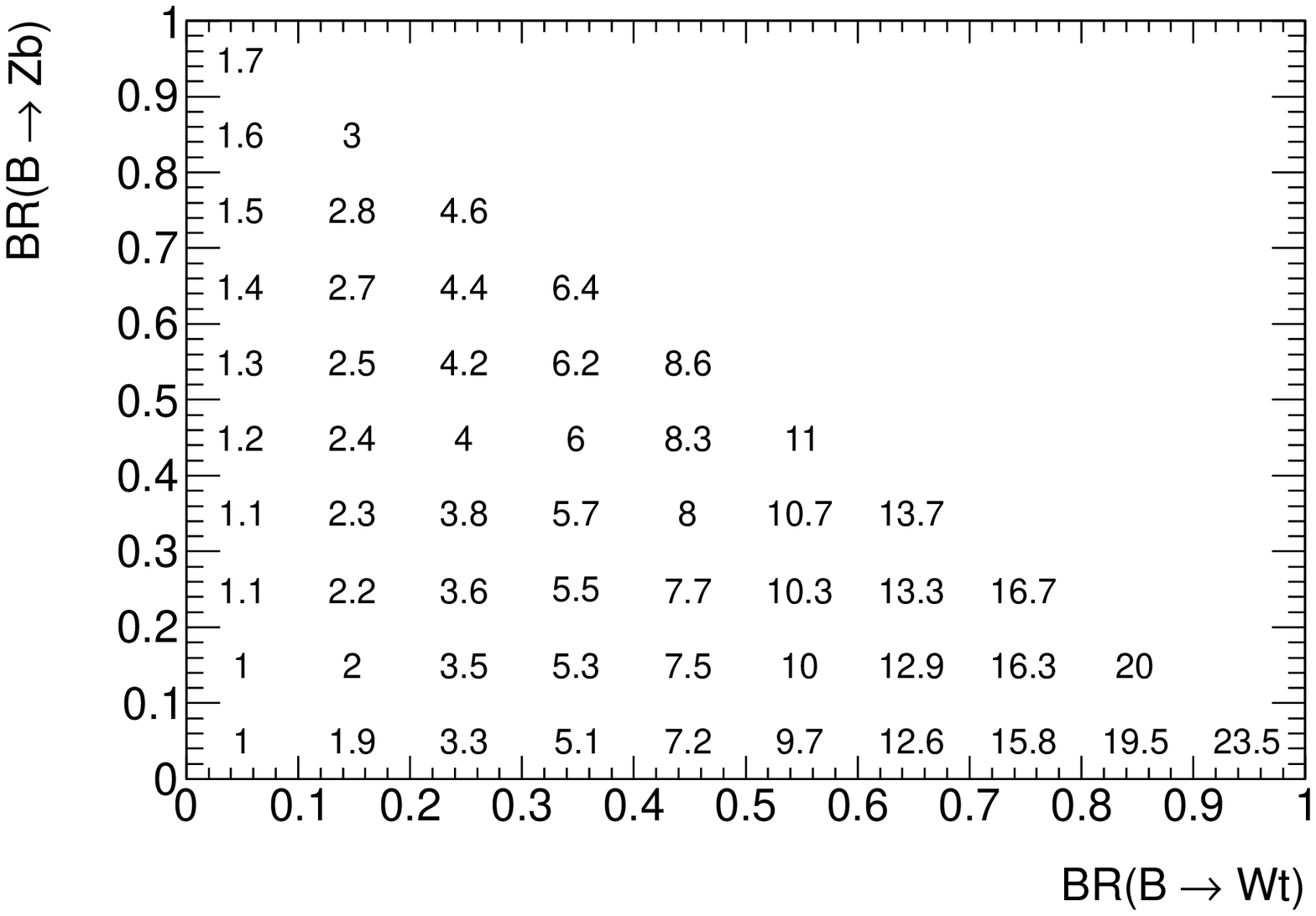}
}
\subfigure[] {
  \includegraphics[height=2.3in]{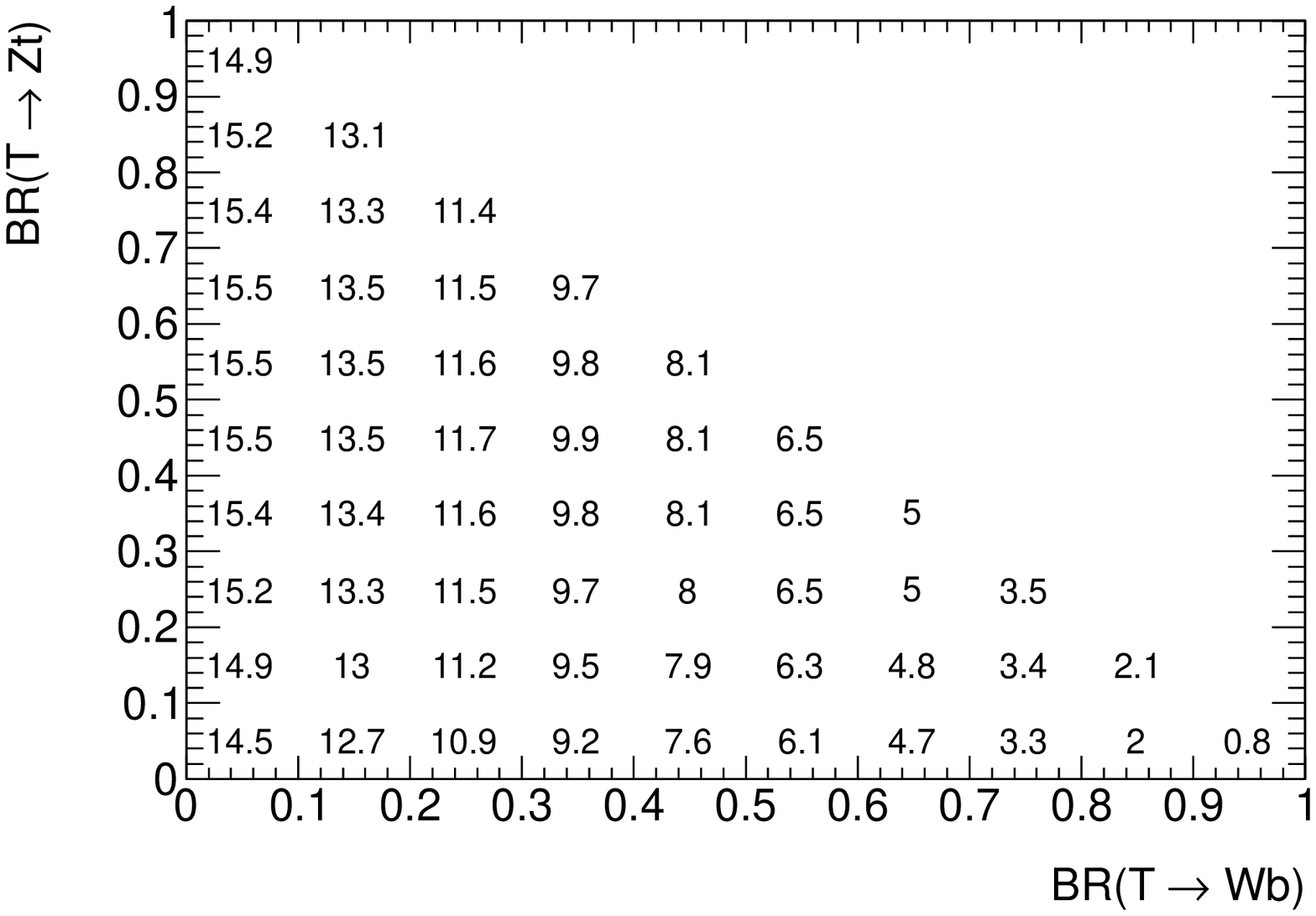}
}
\subfigure[] {
  \includegraphics[height=2.3in]{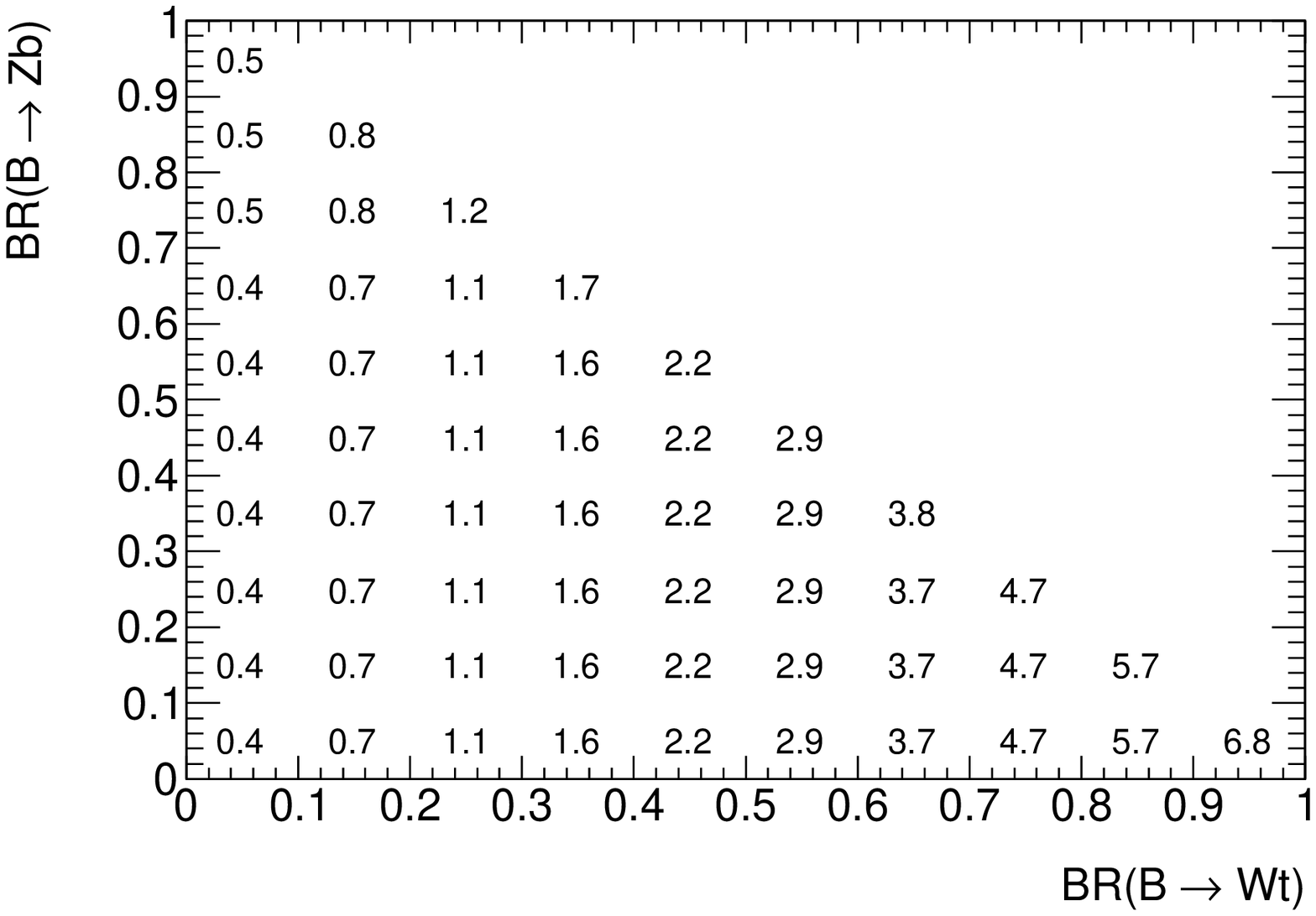}
}
\subfigure[] {
  \includegraphics[height=2.3in]{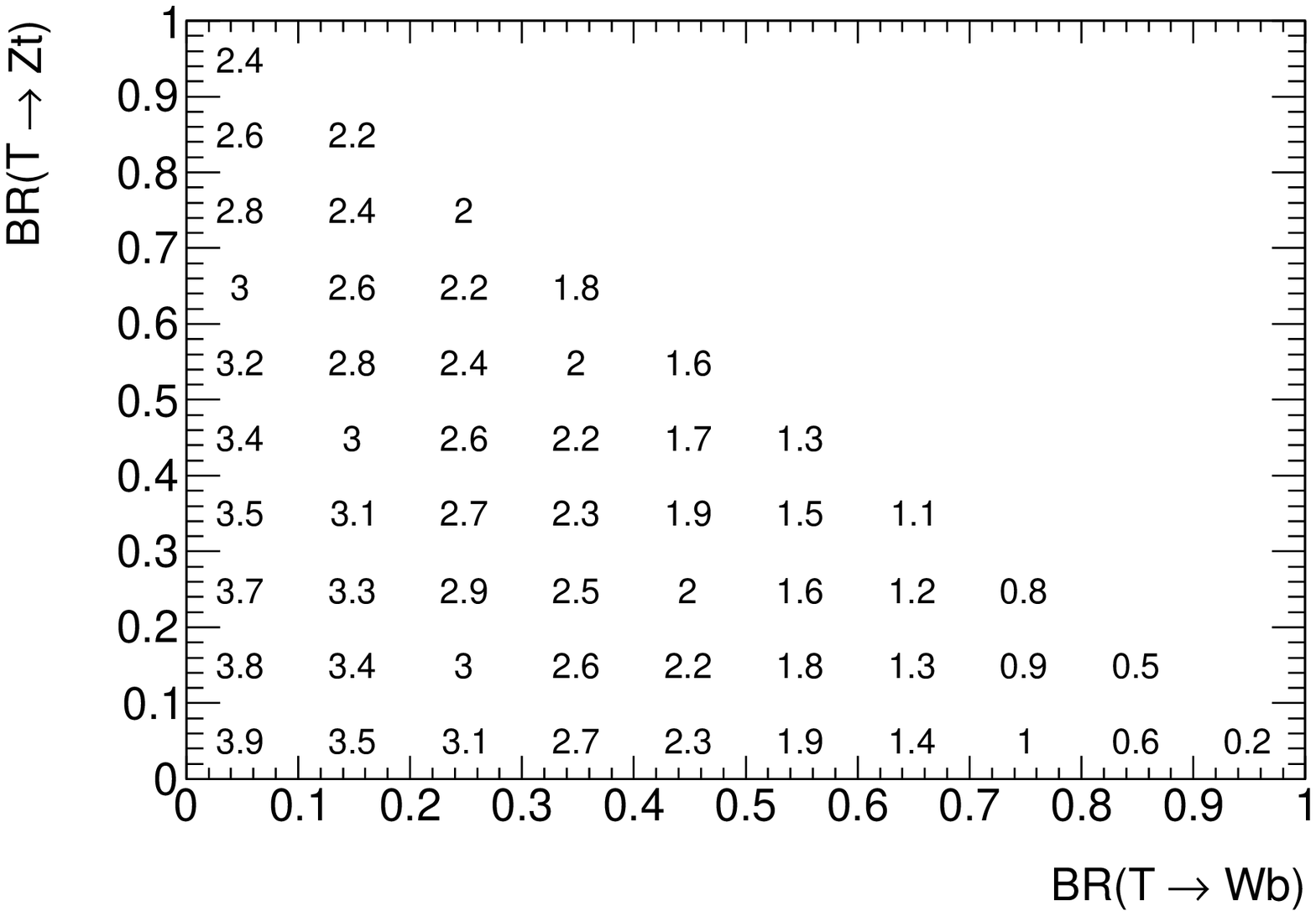}
}
\caption{Here we show the number of expected signal events depending
  on the model for pair production of quarks with a mass of 500~GeV:
  same-sign yields for $B\bar{B}$ (a) and $T\bar{T}$ (b) production,
  and trilepton yields for $B\bar{B}$ (c) and $T\bar{T}$ (d)
  production. }
\label{fig:NEvtAll}
\end{figure*}

After selecting events from the signal samples that pass the selection
requirements, the number of expected events as a function of the
branching ratio of the decays of the new heavy quarks are determined
according to the prescription of Section~\ref{sec:BRConversion}. These
numbers are shown for example production processes in
Figure~\ref{fig:NEvtAll}. In each case the number of events are then
interpreted as an exclusion limit according to the parameterizations
of Figure~\ref{fig:LimitParam}. 

\begin{figure*}
\centering
\subfigure[] {
  \includegraphics[height=2.3in]{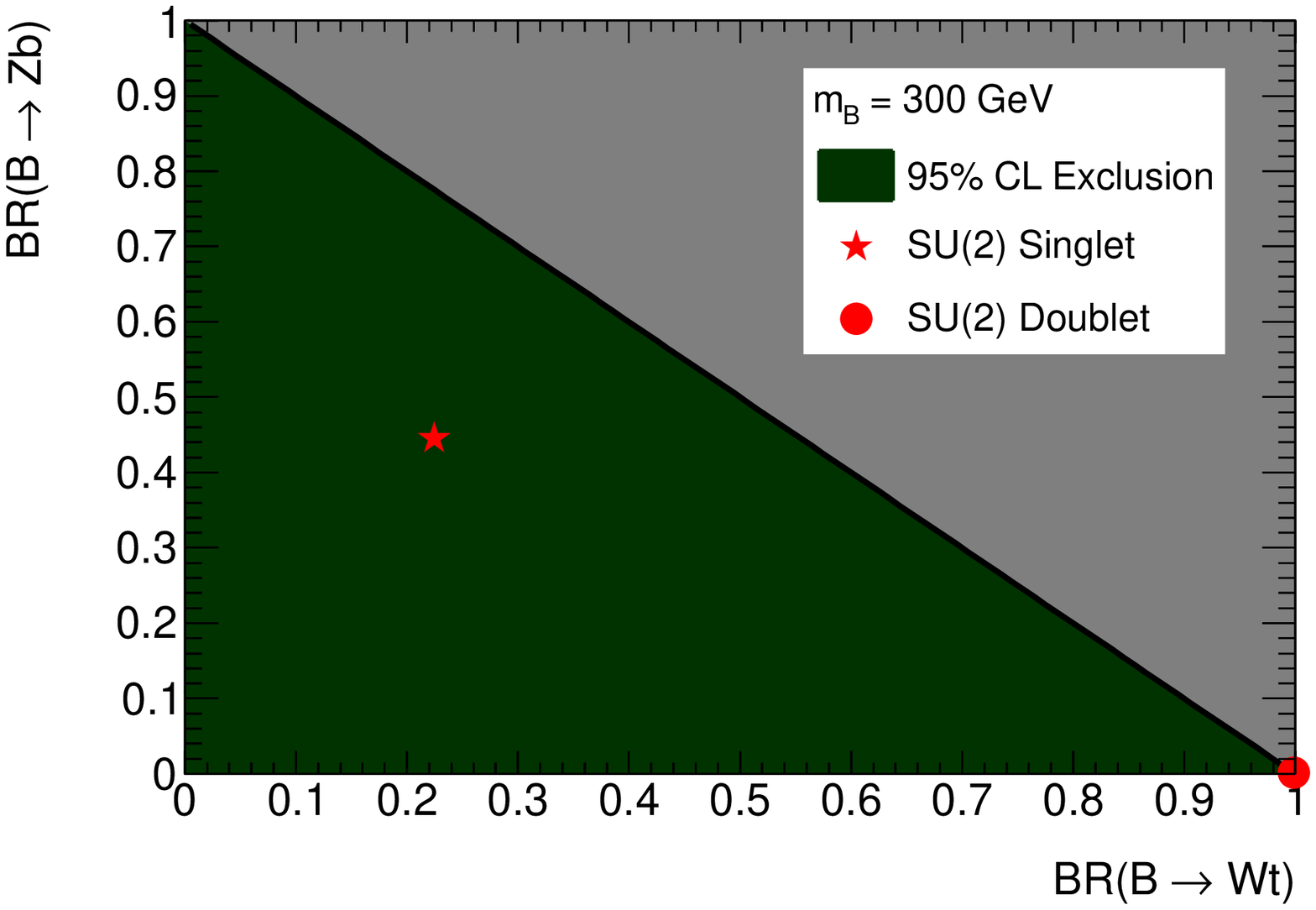}
}
\subfigure[] {
  \includegraphics[height=2.3in]{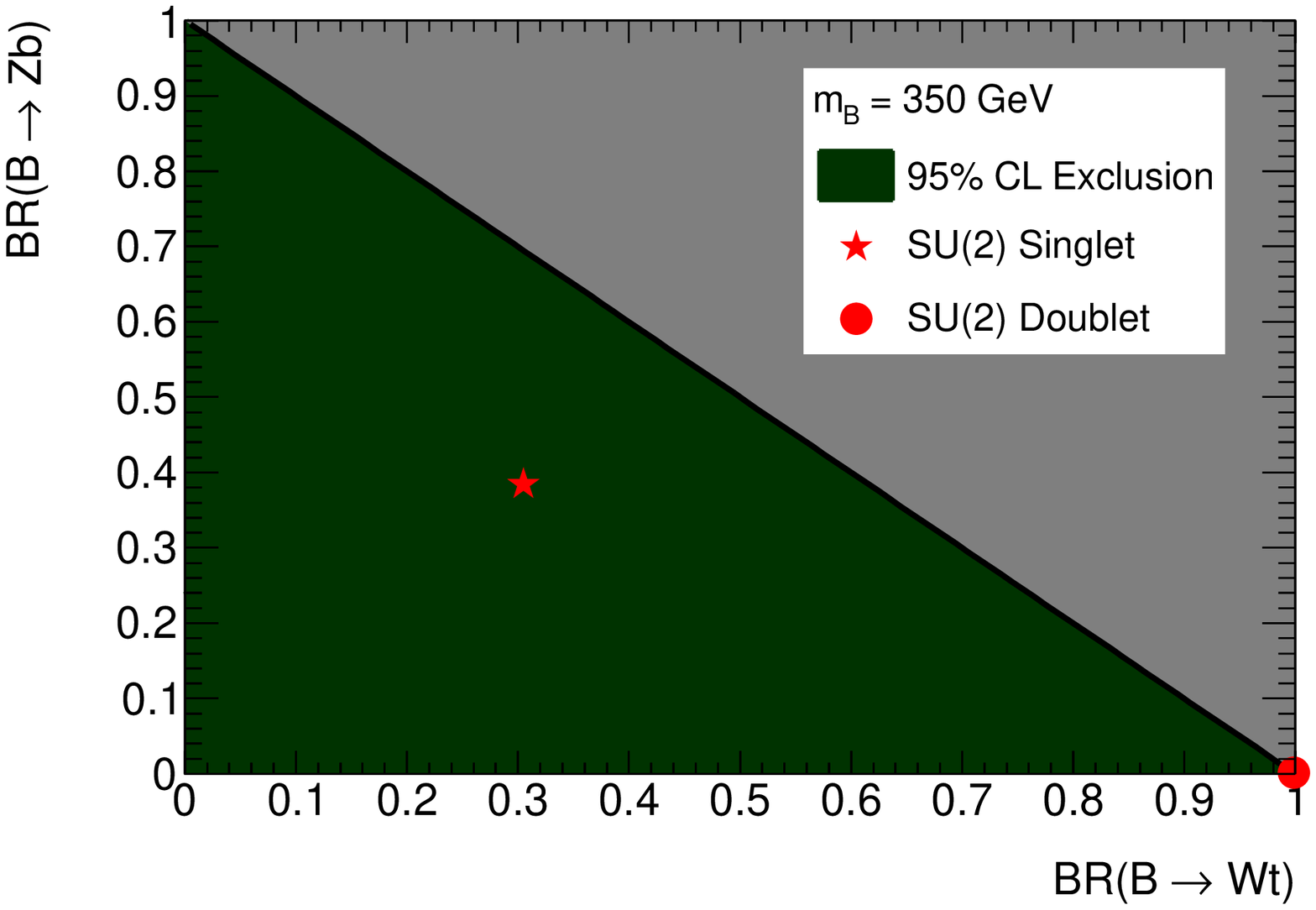}
}
\subfigure[] {
  \includegraphics[height=2.3in]{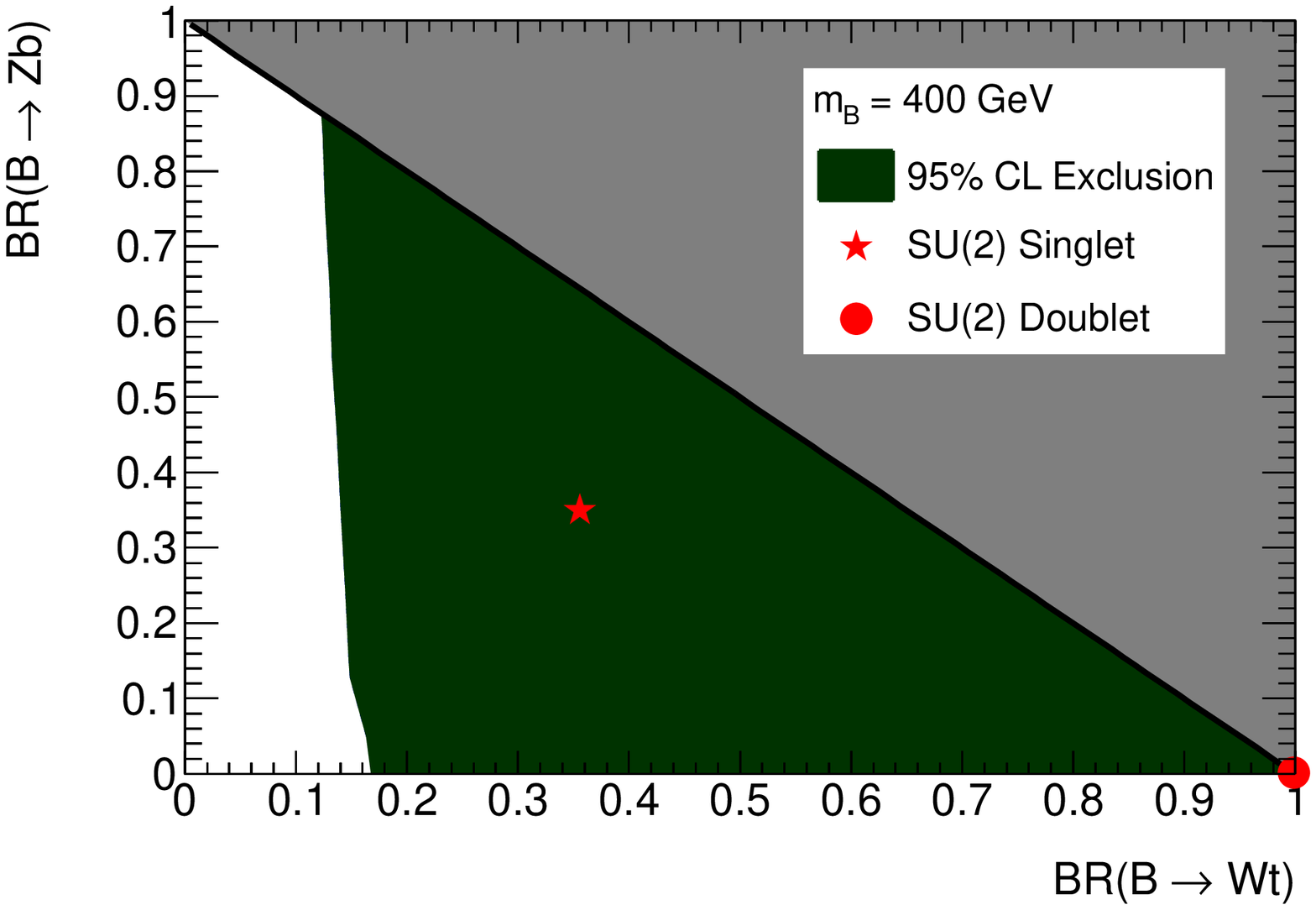}
}
\subfigure[] {
  \includegraphics[height=2.3in]{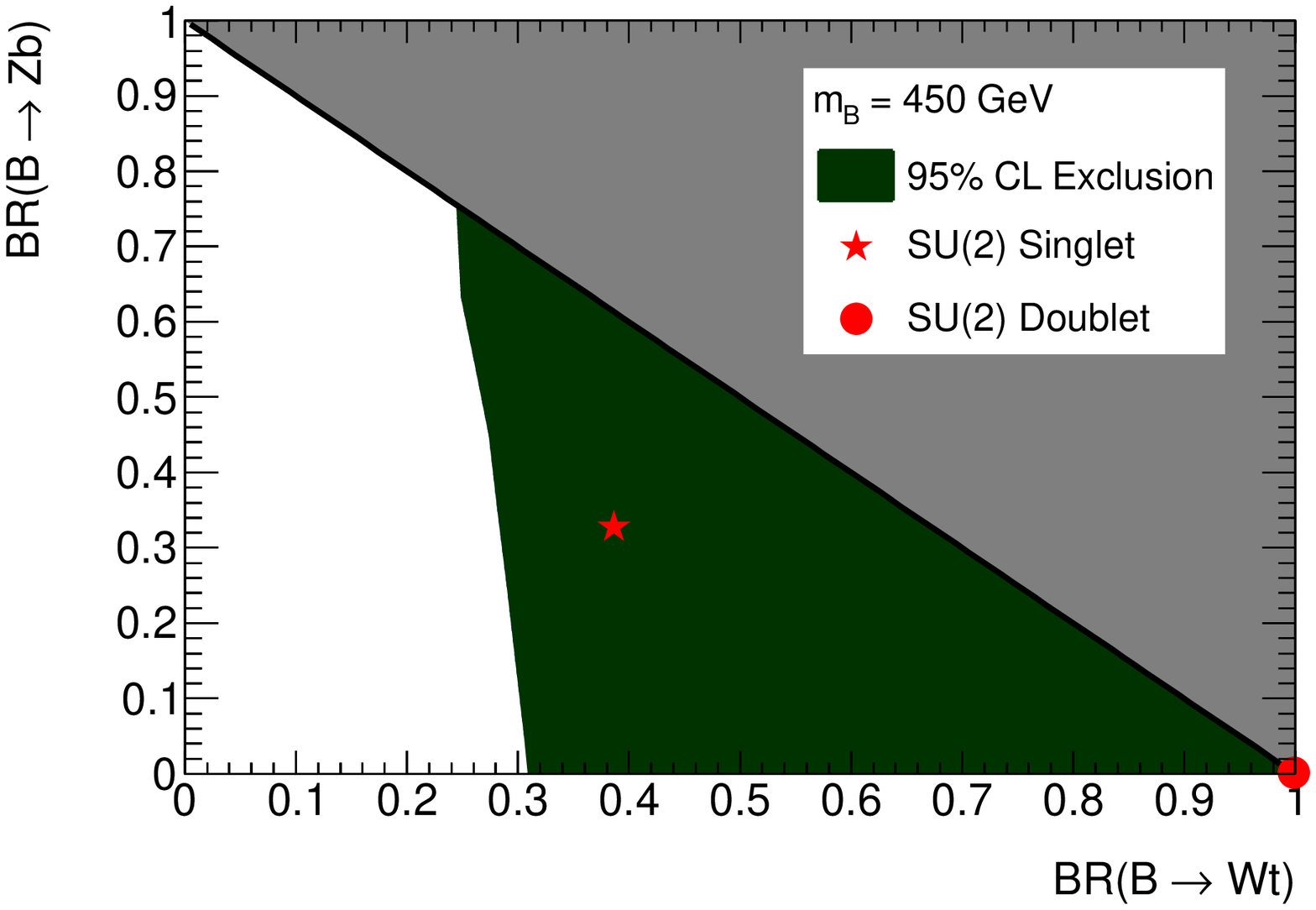}
}
\subfigure[] {
  \includegraphics[height=2.3in]{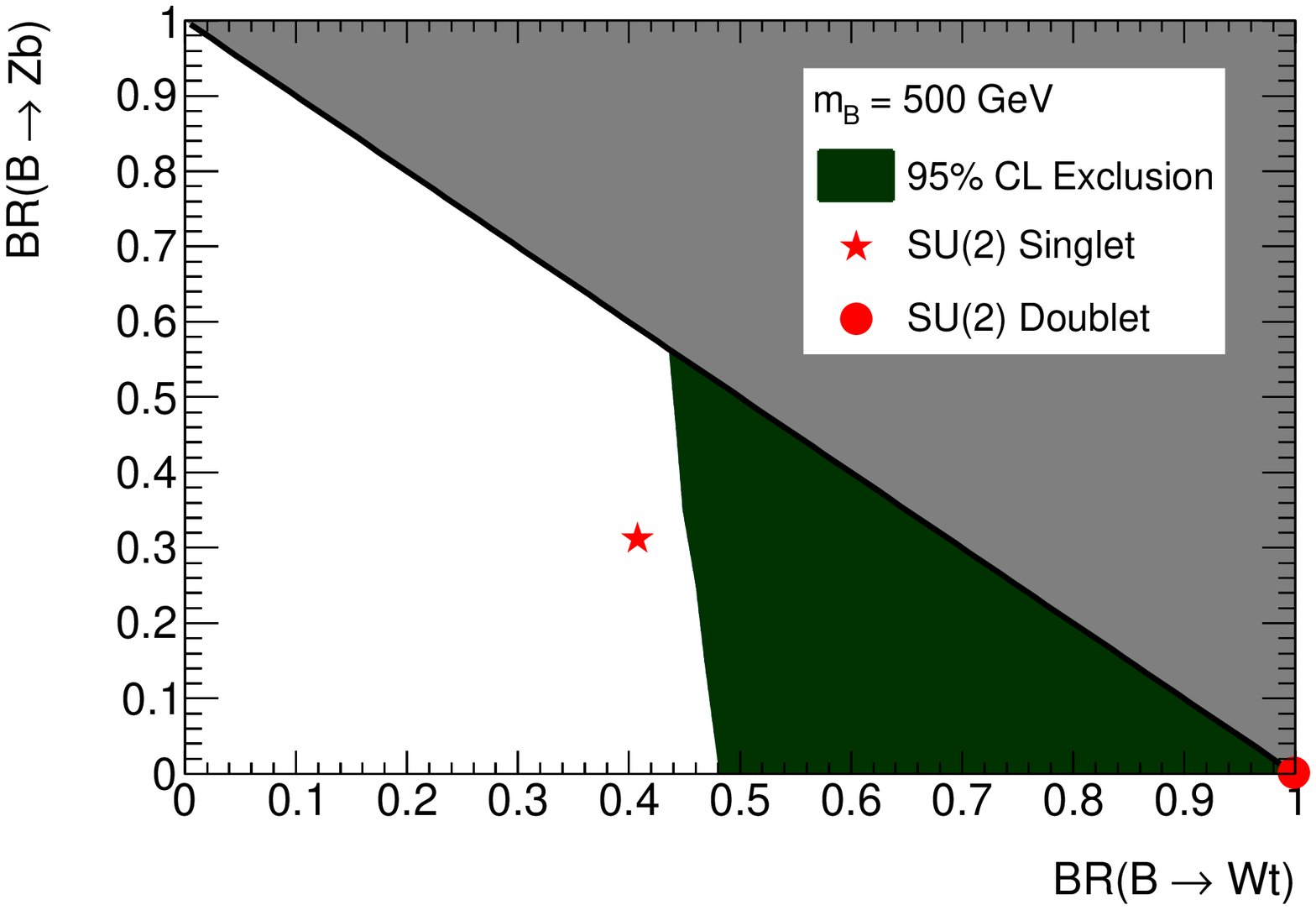}
}
\subfigure[] {
  \includegraphics[height=2.3in]{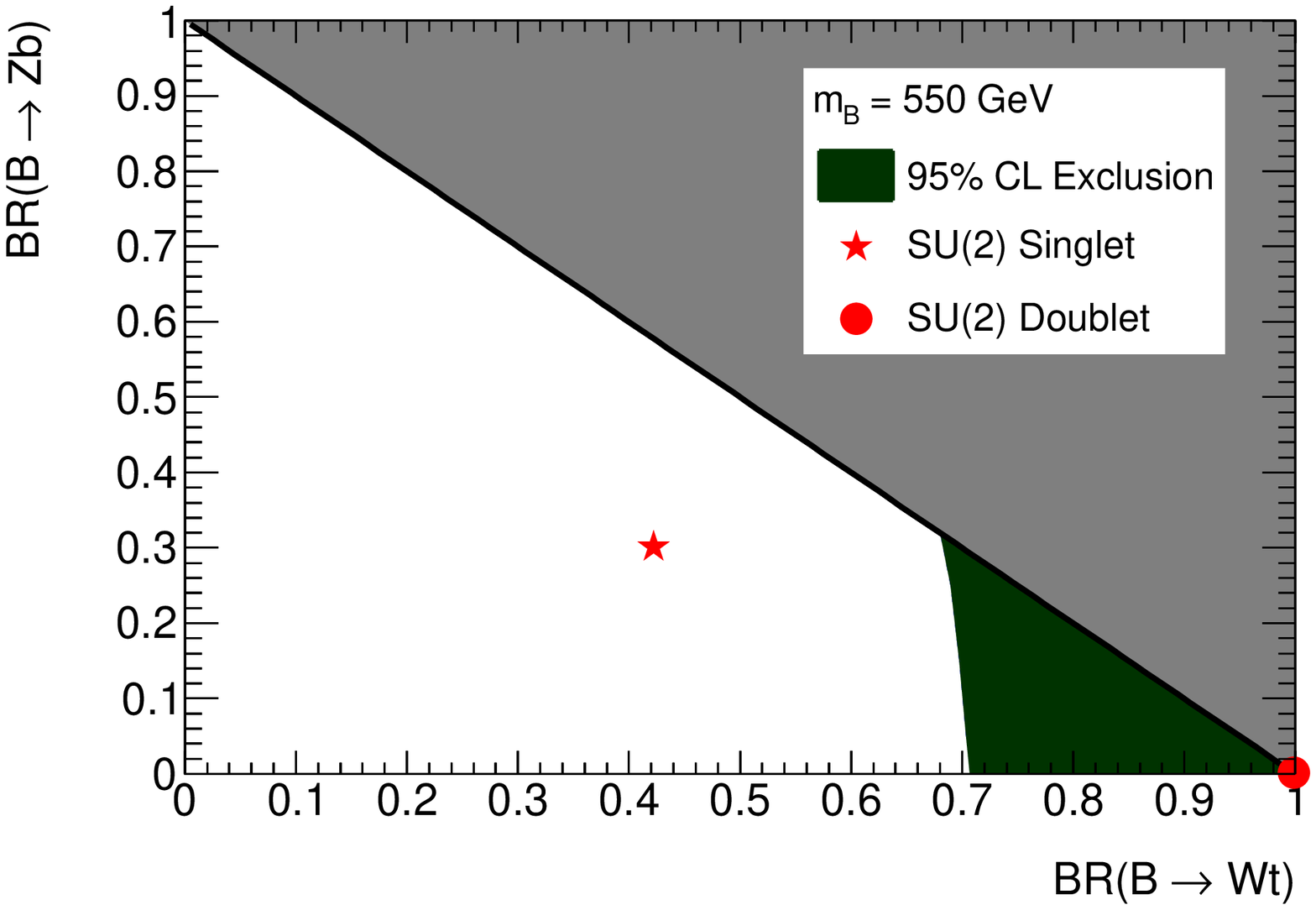}
}
\caption{Here we show the 95\% CL exclusion regions for $B\bar{B}$ pair production
  depending on the assumed branching ratios. The actual branching fractions depend on the
  parameters of the model. In each case the branching fractions of the singlet model 
  are indicated by a star. For the doublet model, the branching ratios depend on the CKM 
  parameters. Branching fractions under the reasonable scenario of $V_{Tb}\ll V_{tB}$ 
  are shown as a circle. 
  Results are shown assuming a $B$ mass of 300~GeV (a), 350~GeV (b),
  400~GeV. (c), 450~GeV (d), 500~GeV (e) and 550~GeV (f). }
\label{fig:LimBB}
\end{figure*}

\begin{figure*}
\centering
\subfigure[] {
  \includegraphics[height=2.3in]{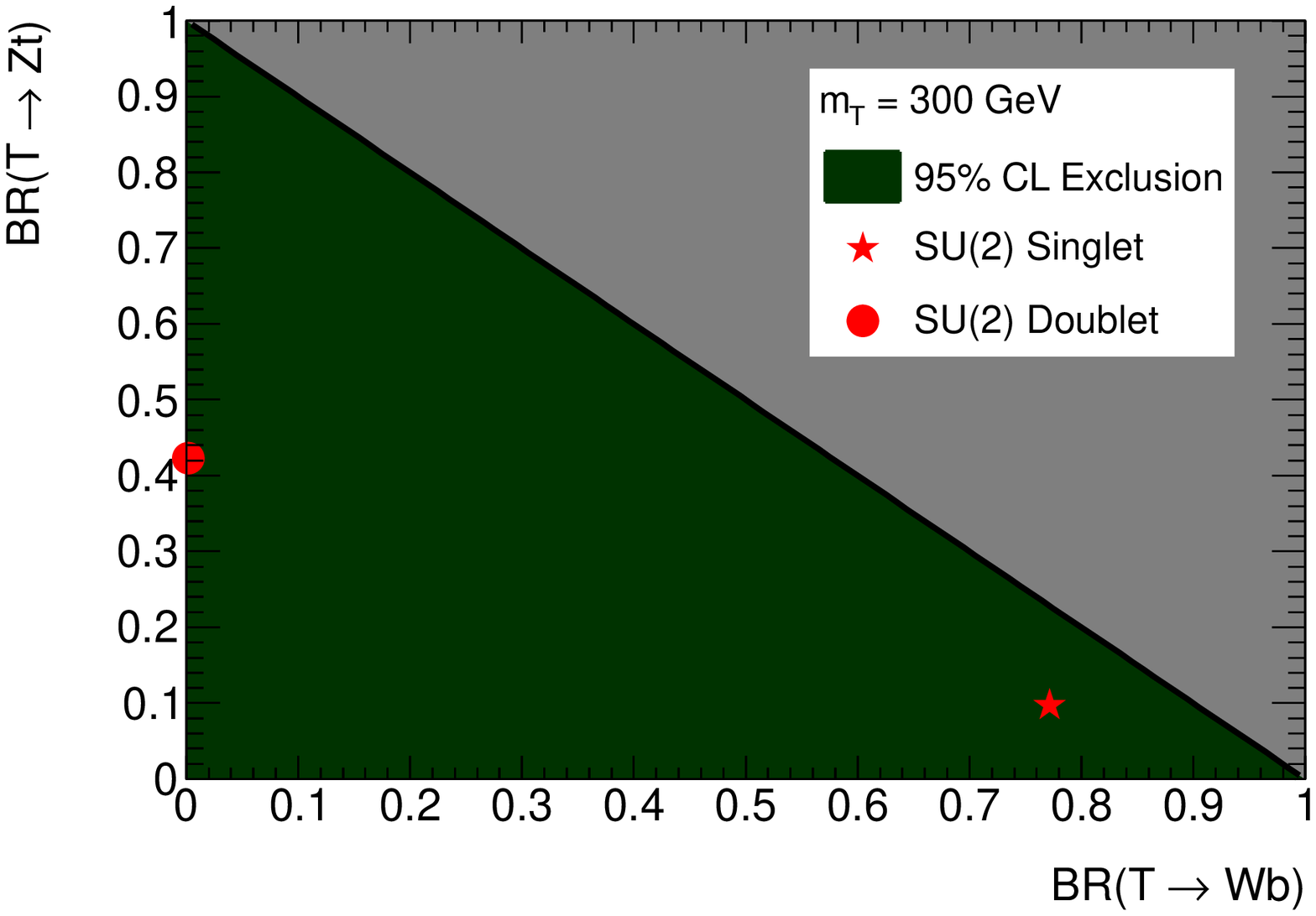}
}
\subfigure[] {
  \includegraphics[height=2.3in]{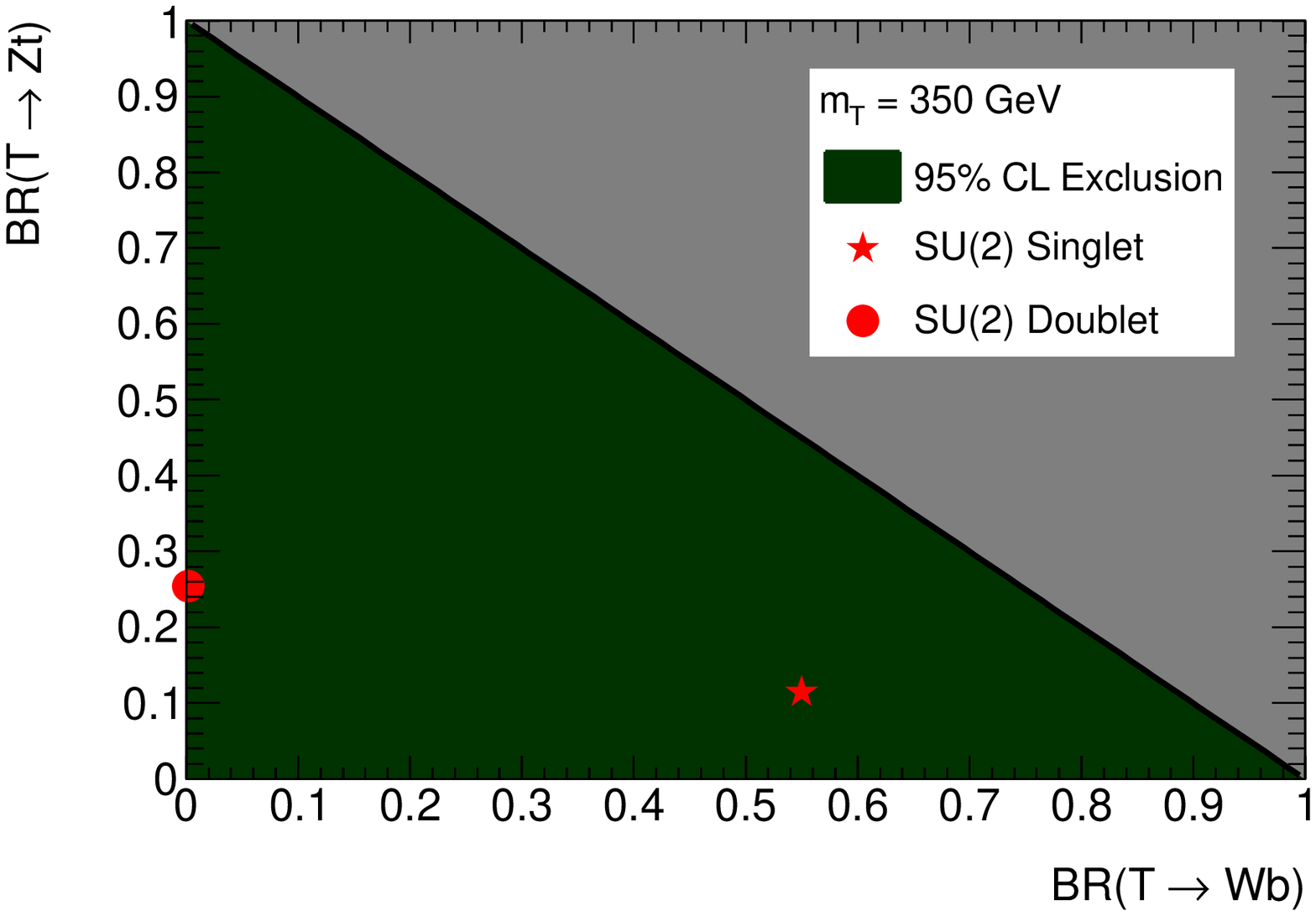}
}
\subfigure[] {
  \includegraphics[height=2.3in]{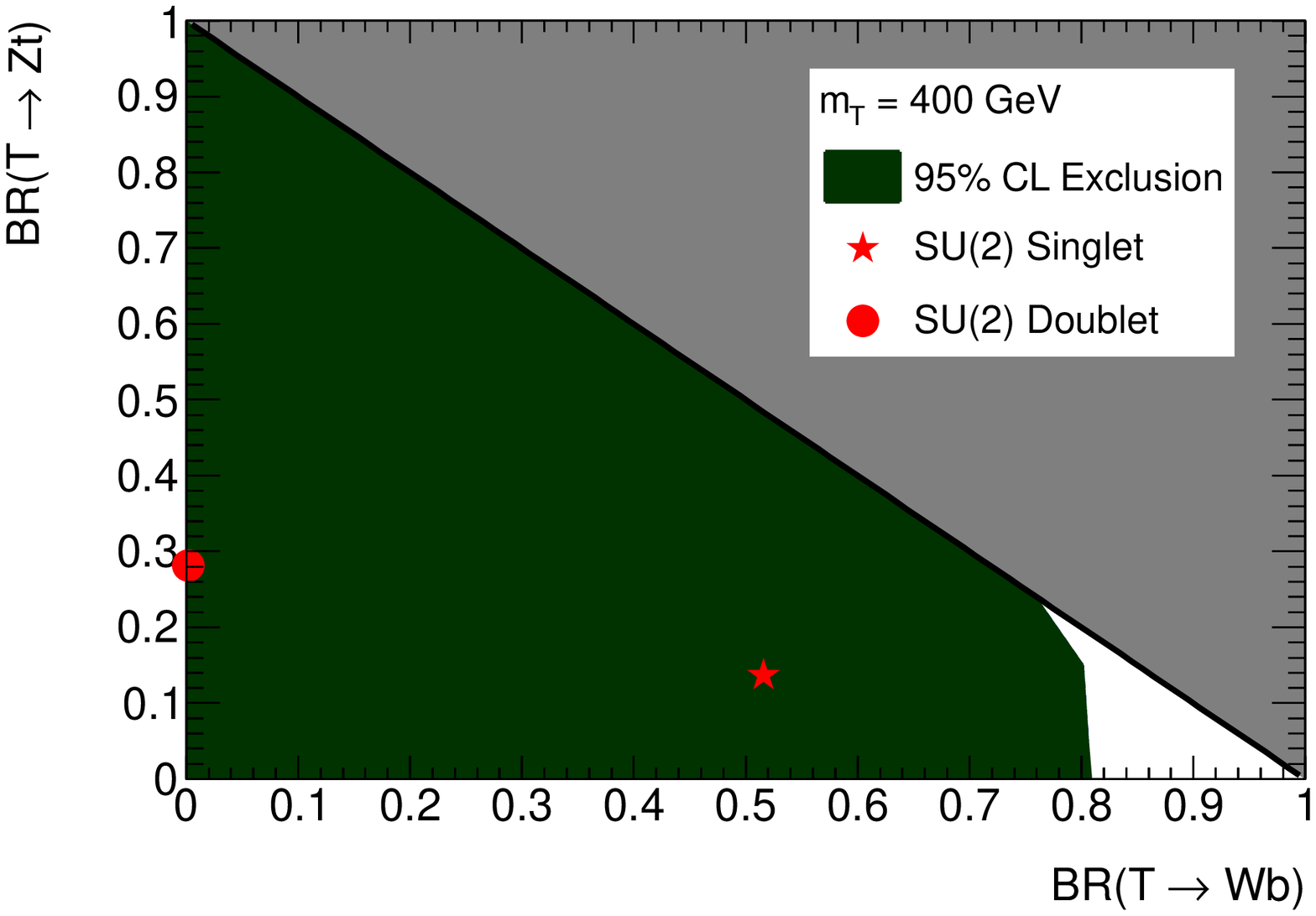}
}
\subfigure[] {
  \includegraphics[height=2.3in]{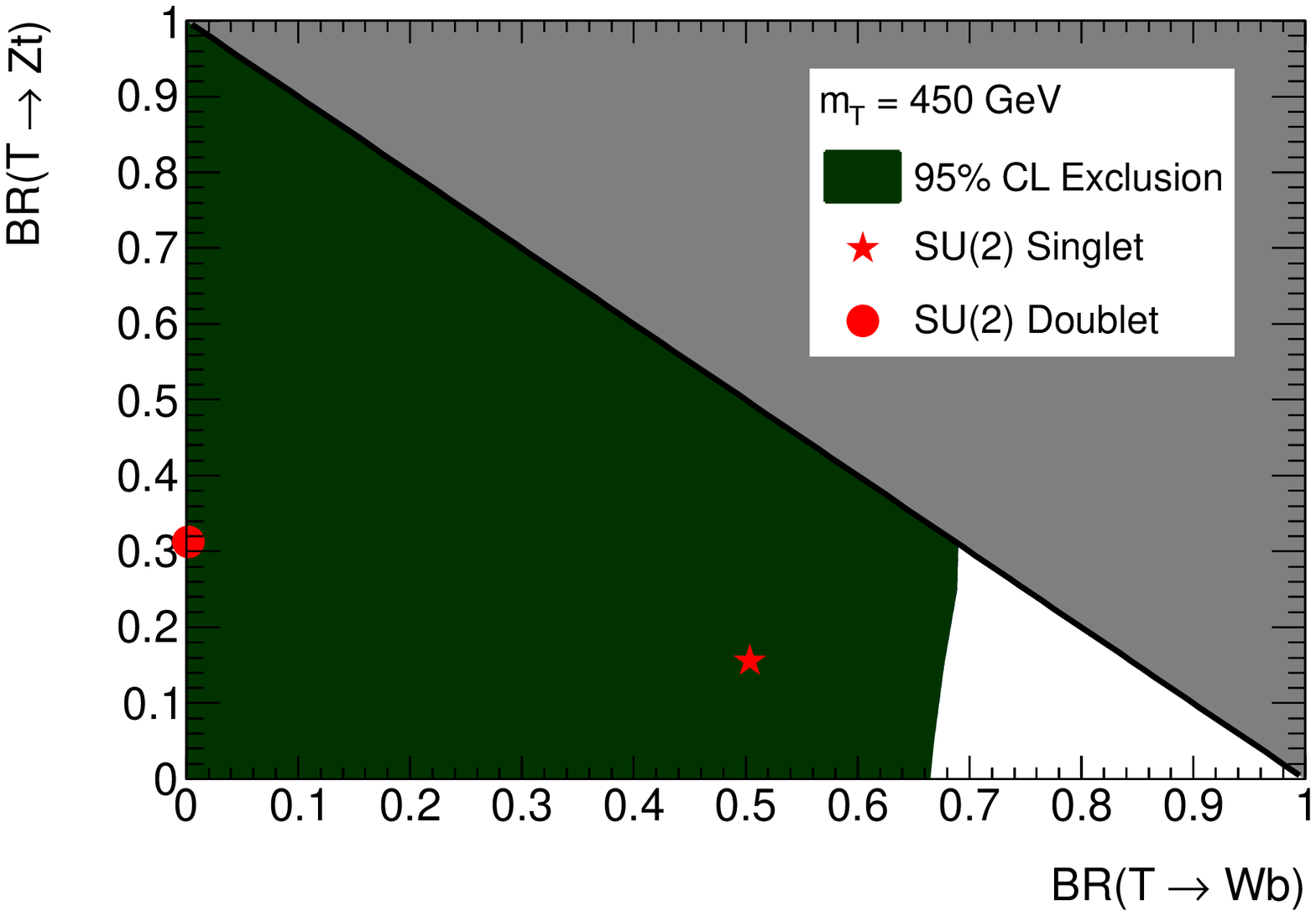}
}
\subfigure[] {
  \includegraphics[height=2.3in]{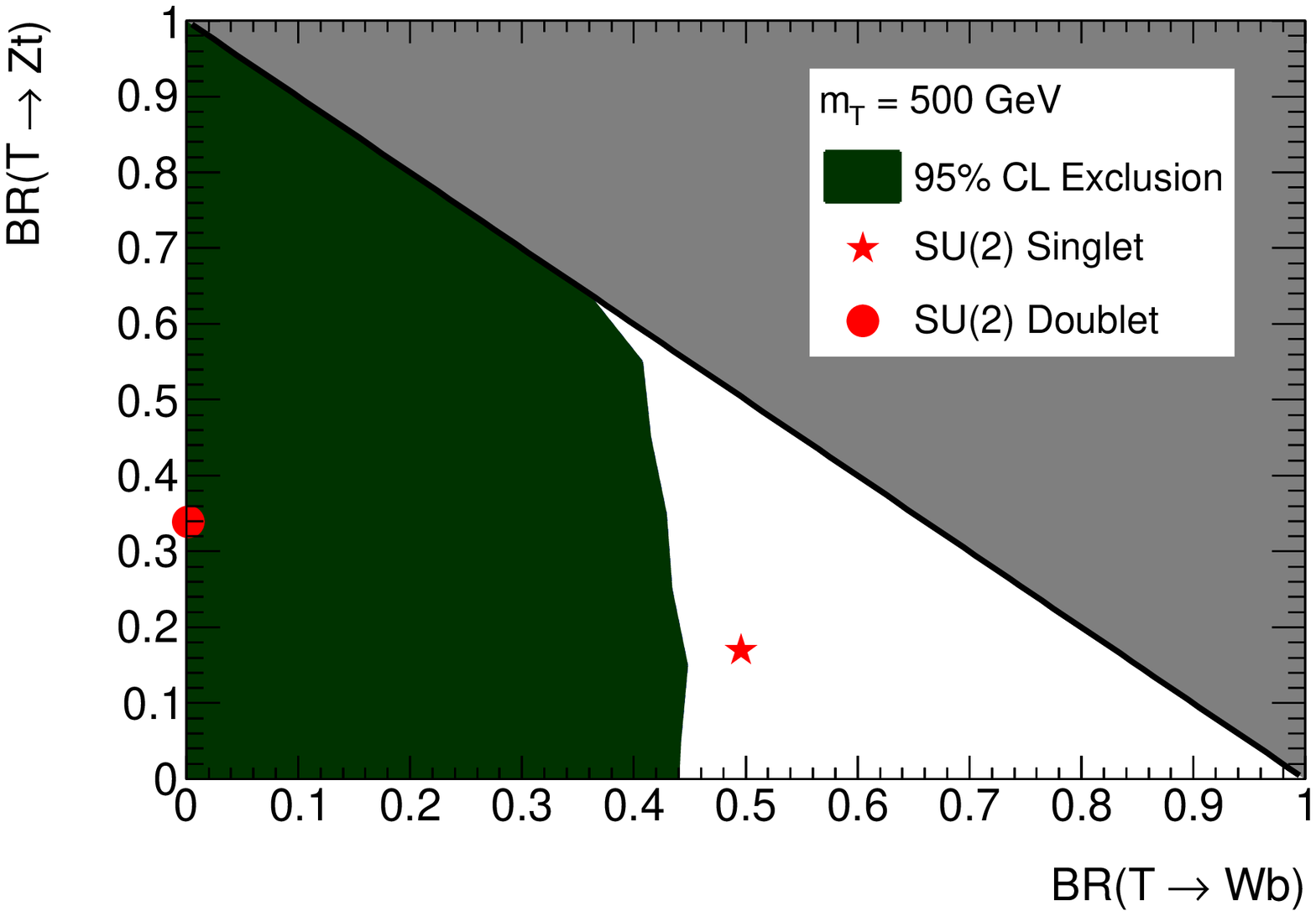}
}
\subfigure[] {
  \includegraphics[height=2.3in]{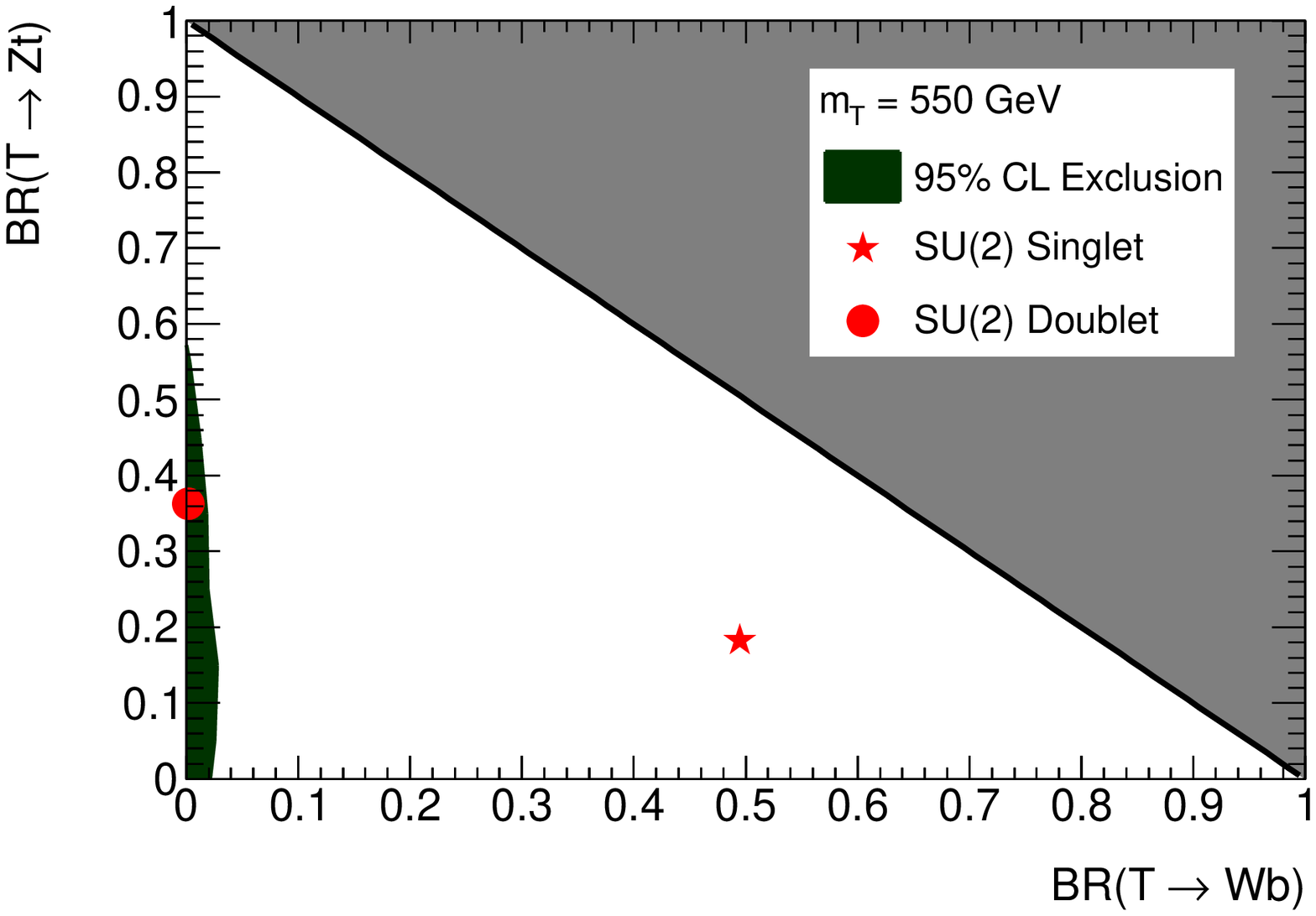}
}
\caption{Here we show the 95\% CL exclusion regions for $T\bar{T}$
  pair production depending on the assumed branching ratios. The actual branching fractions depend on the
    parameters of the model. In each case the branching fractions of the singlet model are indicated by a star.
      For the doublet model, the branching ratios depend on the CKM parameters. Branching fractions under
        the reasonable scenario of $V_{Tb}\ll V_{tB}$ are shown as a circle.
        Results are shown assuming a $T$ mass of 300~GeV (a), 350~GeV (b),
  400~GeV. (c), 450~GeV (d), 500~GeV (e) and 550~GeV (f).}
\label{fig:LimTT}
\end{figure*}

The resulting exclusion values are shown in Figures~\ref{fig:LimBB}
and~\ref{fig:LimTT} for the possible signal hypotheses.  At 95\% CL we
exclude $T$ quarks with a mass $m_T < 480$~GeV and $m_T < 550$~GeV for
weak isospin singlets and doublets, respectively and $B$ quarks with a
mass $m_B < 480$~GeV for singlets.  These doublet-limits are quoted
assuming $V_{Tb}\ll V_{tB}$.  Production of $B\bar{B}$ and $T\bar{T}$
are both excluded at 95\% CL for all possible branching ratios up to a
mass of 360~GeV.  When considering the results of the ATLAS
search~\cite{ATLAS:2012qe} in addition to those of this paper, the
$T\bar{T}$ hypothesis is excluded for all branching ratios up to a
mass of 450~GeV.

\subsection{\label{sec:NominalBR} Results assuming nominal branching fractions for the new heavy quark decays}
In this section results are presented under certain plausible
models. After determining the number of events that are expected to
pass selection for each signal hypothesis, the results are interpreted
as exclusion limits according to the parameterizations of
Figure~\ref{fig:LimitParam}.

\begin{figure*}
\centering
\subfigure[] {
    \includegraphics[height=2.3in]{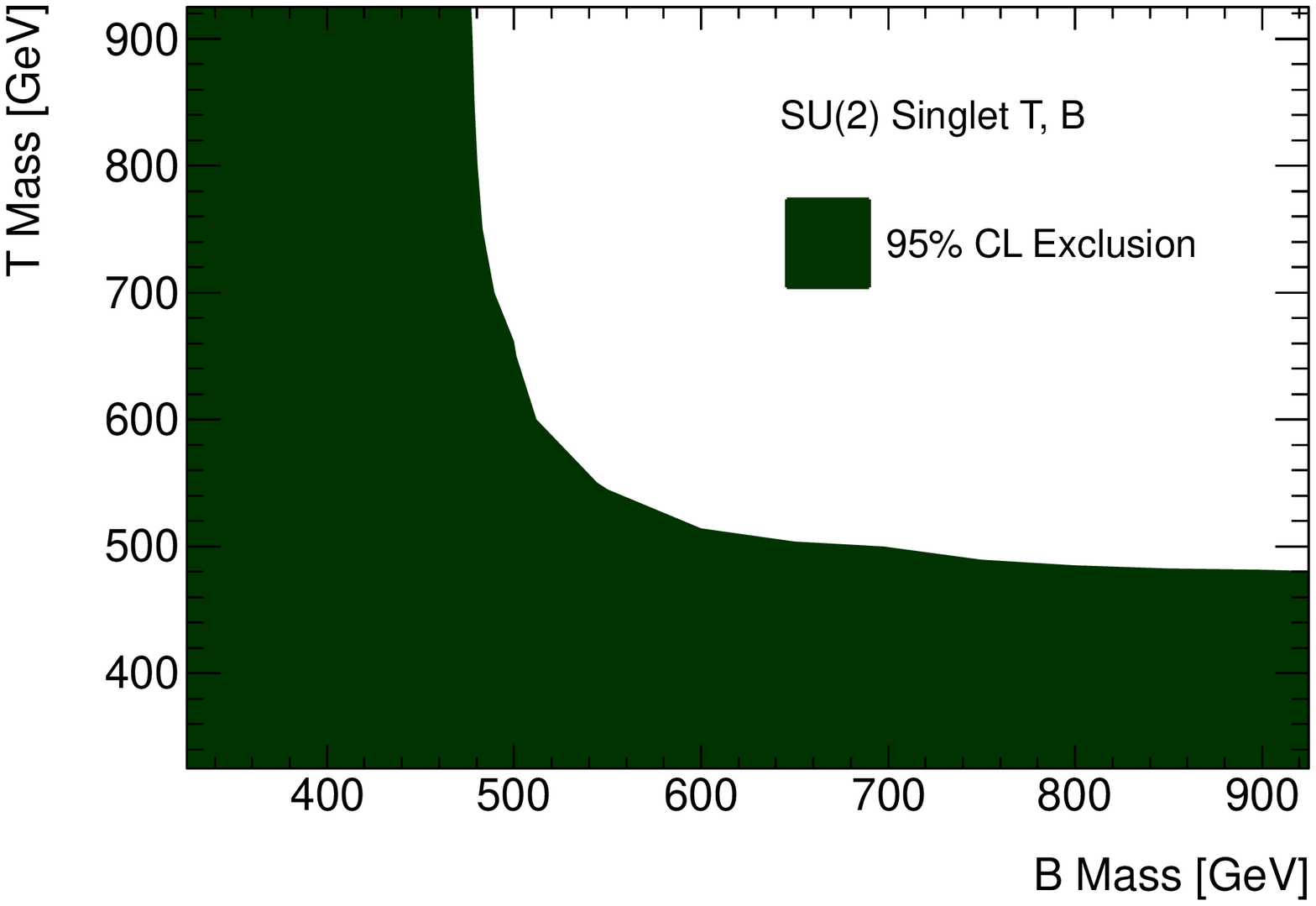}
}
\subfigure[] {
    \includegraphics[height=2.3in]{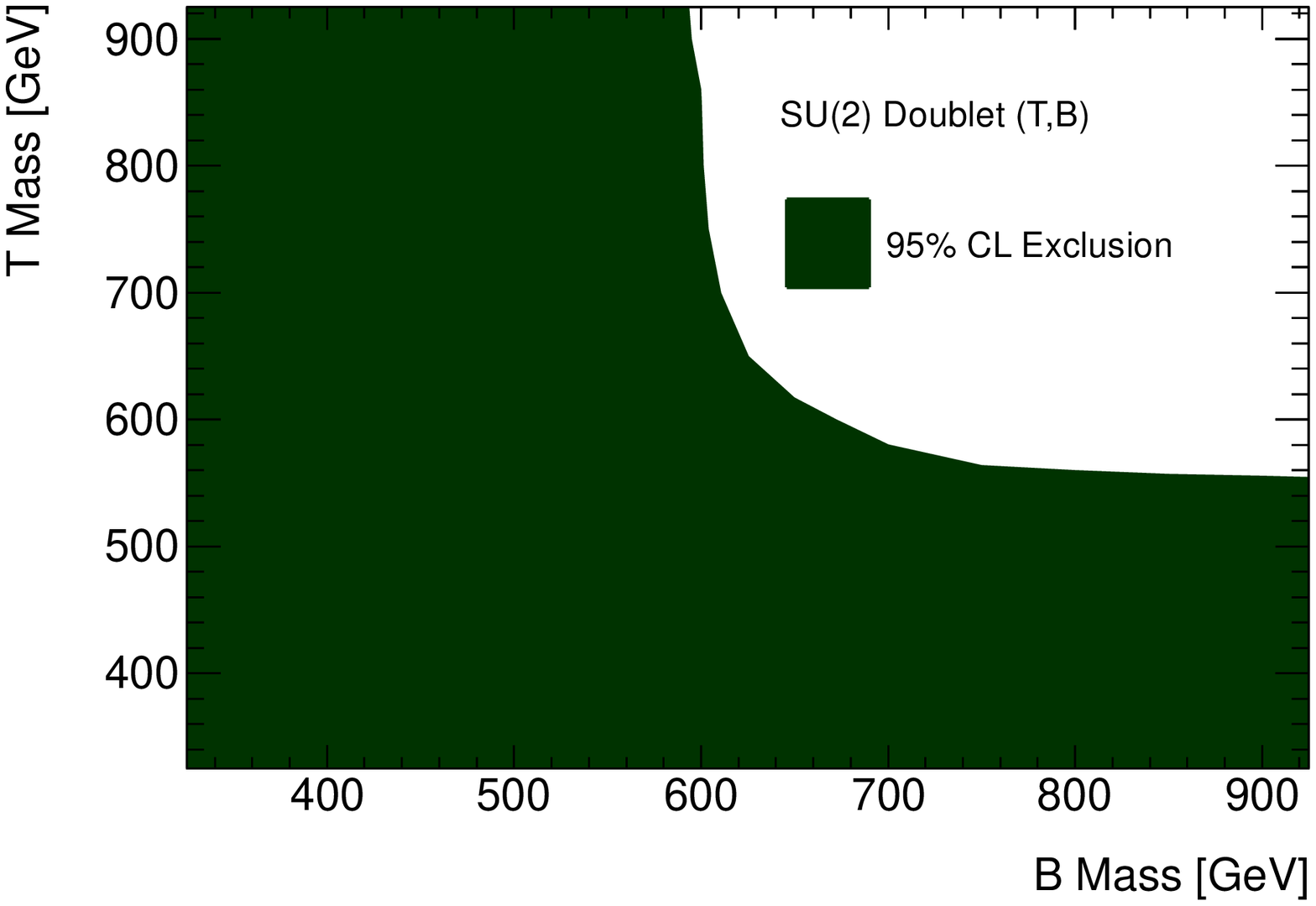}
}
\subfigure[] {
    \includegraphics[height=2.3in]{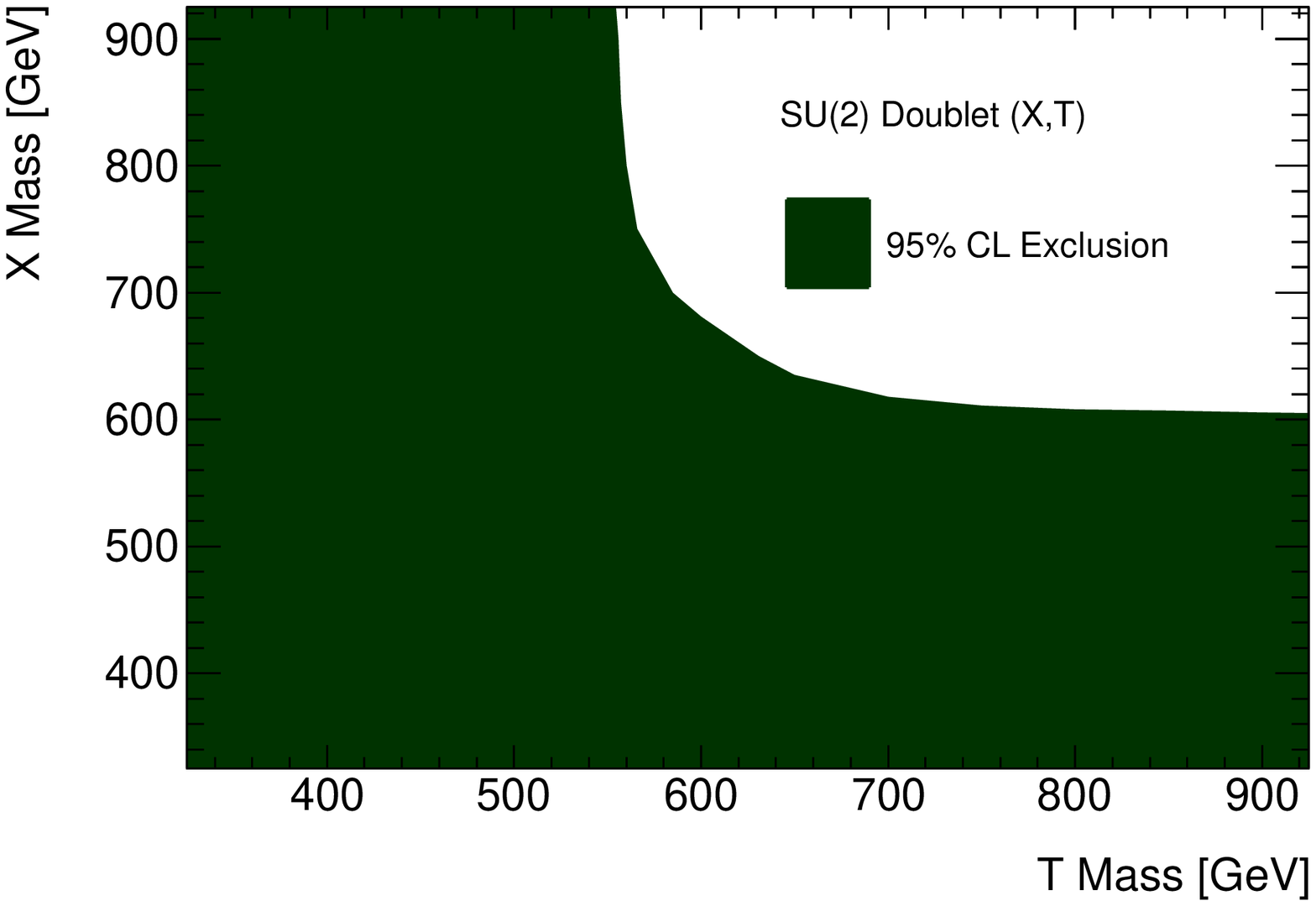}
}
\caption{Here 95\% CL exclusion limits are shown 
  depending on the mass of each quark and the model. Figure (a) shows
  results assuming the presence of two singlets, $T$ and $B$. 
  Figure (b) shows results assuming the presence of a $(T,B)$
  doublet under the assumption $V_{Tb}\ll V_{tB}$. 
Figure
  (c) shows results assuming the presence of a $(X,T)$ doublet.  }
\label{fig:SingletDoubletInclusiveResults}
\end{figure*}

The first hypothesis is that both a $B$ and a $T$ singlet are present
with the expected branching ratios. In this case the default Protos model has
the correct branching fractions and no corrections are required.  
Alternately, we consider the presence of a $(B,T)$ (with $V_{Tb}\ll V_{tB}$) or a $(X,T)$ doublet.
For each of these models, the limit results are presented in
Figure~\ref{fig:SingletDoubletInclusiveResults} in a two-dimensional
grid depending on the hypothesized mass of the new heavy quarks. It should
be noted that in the case of the singlet model there is no reason to assume that
both a $B$ and a $T$ quark must be present. In case only a single quark is present,
the limits can be extracted by considering the high-mass limit for the other
quark in Figure~\ref{fig:SingletDoubletInclusiveResults}.

Assuming identical masses for the new heavy quarks in the singlets
and doublets, the 95\% CL limits can be interpreted as $m_{Q} <
550$ GeV for the case of a singlet $T$ and a singlet $B$, $m_{Q} <
640$ GeV for the case of a doublet $(T,B)$, and $m_{Q} < 640$ GeV for
the case of a doublet $(X,T)$.

\section{\label{sec:conclusion}Conclusion}

We demonstrate in this paper that searches in final states with three
isolated leptons ($e$ or $\mu$) or two isolated leptons with same
electric charge have a very good sensitivity to exotic heavy quarks
$T$, $B$ and $X$ for all possible decay modes $T\to W^+b$, $T\to Zt$,
$T\to Ht$, $ B\to W^-t$, $B\to Zb$, $B\to Hb$ and $X\to W^+t$.  ATLAS
and CMS searches in these final states have previously set limits
assuming $BR(b' \to W^-t) = 1$ and $BR(X \to W^+t) = 1$.  We
reinterpret CMS results and generalize their limits for arbitrary
branching ratios for heavy quark masses above 300~GeV.

For vector-like quark models we exclude at 95\% CL $T$ quarks with a
mass $m_T < 480$~GeV and $m_T < 550$~GeV for weak isospin singlets and
doublets, respectively and $B$ quarks with a mass $m_B < 480$~GeV for
singlets.  Mass limits at 95\% CL for $T$ and $B$ singlets, ($T$,$B$)
doublets and ($X$,$T$) doublets are presented as a function of the
corresponding heavy quark masses.  Under an equal mass hypothesis
($m_T = m_B$ and $m_X = m_T$) vector-like quarks are excluded at 95\%
CL with masses below 550~GeV for $T$ and $B$ singlets, 640~GeV for
($T$,$B$) doublets (assuming $V_{Tb}\ll V_{tB}$) and 640~GeV for
($X$,$T$) doublets.

\section{\label{sec:acknowledgements}Acknowledgements}
We thank Juan Antonio Aguilar-Saavedra and Nuno Castro for useful
comments. The authors are supported by grants from the Department of
Energy Office of Science and by the Alfred P. Sloan Foundation.

\bibliography{reinterpretation_outline}
\bibliographystyle{ieeetr}

\end{document}